\documentclass[twocolumn,showpacs,preprintnumbers,nofootinbib]{revtex4}

\usepackage{amsmath}
\usepackage{amsfonts}
\usepackage{graphicx}
\usepackage{amssymb,latexsym}
\usepackage{hyperref}

\begin{document}
\title{$\phi^4$ inflation is not excluded}

\author{Erandy Ramirez}
\email{eramirez@physik.uni-bielefeld.de}

\author{Dominik J. Schwarz}
\email{dschwarz@physik.uni-bielefeld.de}

\affiliation{Fakult\"at f\"ur Physik, Universit\"at Bielefeld, 
Universit\"atstra\ss e 25, Bielefeld 33615, Germany}

\begin{abstract} 
We present counterexamples to the claim that the $\lambda \phi^4$ inflaton 
potential is excluded by recent cosmological data. 
Finding counterexamples requires that the actually observed primordial 
fluctuations are generated at the onset of the slow-roll regime of inflation. 
This set up for the initial conditions is therefore different from the usual 
scenario of chaotic inflation where inflation starts long before the 
observed fluctuations are created. The primordial power spectrum 
of ``just enough'' chaotic inflation violates scale-invariance in a 
way consistent with observations.
\end{abstract}

\date{\today}
\preprint{}
\maketitle

\section{introduction and basic idea}

Cosmological inflation is considered to be the essential mechanism for
setting the initial conditions of the standard cosmological scenario. At the
same time it provides an explanation for the formation of structures in
the Universe and there exists a collection of inflationary models whose
predictions can be contrasted to observations. Constraining the 
space of allowed models of inflation is a major goal of CMB experiments 
such as WMAP and Planck.

One of the simplest and most popular models of inflation introduces 
just one self-interacting (real) scalar field $\phi$ in the context of the
chaotic scenario \cite{linde83}. In this scenario the fluctuations 
of matter and space-time observed by means of the CMB temperature 
anisotropies are created long after the onset of inflation. 
Observable modes cross out of the Hubble horizon at 50 to 60 e-foldings 
before the end of inflation, whereas the total duration of chaotic inflation 
is ${\cal O}(10^3)$ e-foldings. Thus the kinetic energy of 
the field plays no significant role for the observable fluctuations  
and the Universe is said to be well within the slow-roll regime. For the 
dynamical evolution of a particular model assumed to be in this stage, it is
justified to apply the slow-roll approximation to the equations of motion 
and therefore to neglect the kinetic energy of the inflaton. It is within this
picture that the $\lambda\phi^4$ potential for inflation has been
excluded by recent observations \cite{wmap5}, although  
including a "curvaton" field relaxes the observational constraints on
this potential \cite{Moroi}.

Here we consider a different situation in which the total amount of inflation
is not much more than 60 e-foldings. The onset of inflation is 
thus observable and therefore the effect of the kinetic 
energy is important. The resulting primordial power spectrum is not scale 
invariant, as the moment of the onset of the slow-roll regime distinguishes 
a scale. We find that a negative running of the spectral index is a feature 
of this scenario. This also means that the Universe undergoes only a small 
amount of inflation before entering the slow-roll regime.
Although not in accordance with the generic initial conditions of chaotic 
inflation, such a situation cannot be discarded on grounds of current 
analysis and observations. 

This new scenario of ``just enough'' chaotic inflation seems to be generic 
if two fundamental energy scales are relevant in the very early Universe. The
Planck scale $M_p$ is the fundamental scale of quantum gravity. The notions 
of spatial 
curvature, expansion rate $H$ and kinetic energy density 
$\frac 12 \dot{\phi}^2$ seem 
to be well defined and real quantities, at least up to that scale. 
However, this is less clear for the effective potential $V$ of the inflaton. 
The effective potential carries all information about all the interactions 
of the inflaton except its gravitational ones. There exist examples 
of (low-energy) effective field theories whose potential becomes complex
or show a singularity at some high-energy scale that is still well below the 
Planck scale. 

The standard model of particle physics is one of these examples:
The quartic self-coupling of the Higgs runs with energy. Except for a 
very heavy Higgs, the self-coupling decreases with increasing energy scale 
and can even become negative at high energy. The effective Higgs potential 
is real as long as the quartic coupling is positive, but becomes complex 
as soon as the self-coupling runs to negative values.
Thus the standard model provides an example in which loop corrections 
give rise to an imaginary contribution of the effective potential. Such an 
imaginary contribution would give rise to the decay of the 
Higgs field. \cite{fjse}.

A fundamental energy scale $M$ acting as an upper bound on the 
effective inflaton potential
also shows up in supergravity models. The effective potential becomes 
too steep for inflation at some high energy scale. Although this 
mechanism gives rise to a low-energy potential that 
does not directly apply to the case we are considering here, it is 
an example of an upper bound for the validity of the inflaton 
potential \cite{kallosh}.

Motivated by these examples, here we consider the possibility that 
there exists a second 
fundamental scale $M < M_p$ that sets an upper bound for the validity 
of a description 
in terms of an effective 
potential $V < M^4$. Above this scale, the potential might be complex, 
giving rise to the 
decay of the inflaton. In the 
context of chaotic inflation 
the inflaton field takes values well above the Planck scale, which is also 
the case in our 
scenario, nevertheless this will not prevent the potential to be of the order 
of $M^4$ at the onset of inflation and below during its evolution. Thus 
just enough chaotic 
inflation follows generically from assuming that the Heisenberg uncertainty 
provides us with 
the initial conditions $H \sim M_p$, $\dot{\phi} \sim M_p^2$, as in standard 
chaotic inflation, 
but $V \sim M^4 \ll M_p^4$ as there exists a fundamental scale $M < M_p$ 
above which the 
inflaton is unstable. As will be 
shown below, the current observational constraints 
are consistent with the assumption that this scale corresponds to that of
grand unification theories $M \sim M_{\rm GUT}$.

\section{$\phi^4$ inflation}\label{chi}

We assume the homogeneity, isotropy and flatness of the Universe from the 
onset of 
inflation, although this is certainly a very rough approximation only. The 
equations of motion 
are then
\begingroup
\everymath{\scriptstyle}
\small
\begin{eqnarray}
\label{ce}
H^2&=&\frac{1}{3M_p^2}\left(\frac{\dot{\phi}^2}{2}+V\right), \\ \nonumber
\ddot{\phi}+3H\dot{\phi}+V'&=&0;
\end{eqnarray}
\endgroup
a prime denotes a derivative with respect to the field,
 $M_p \equiv m_{p}/\sqrt{8\pi} 
\approx 2.4 \times 10^{18}$ GeV is the reduced Planck mass. For the numerical 
analysis we rewrite them as  
\begingroup
\everymath{\scriptstyle}
\small
\begin{eqnarray}
\label{cde}
\frac{dH}{d\phi}&=&\frac{1}{M_p^2}\sqrt{\frac{3M_p^2H^2-V}{2}}, \\ \nonumber
\frac{dN}{d\phi}&=&-\frac{H}{\sqrt{2\left(3M_p^2H^2-V\right)}}.
\end{eqnarray}
\endgroup
Our convention is $\dot{\phi}<0\Rightarrow H'>0$ and
$H \equiv dN/dt$ with $t$ cosmic time, therefore $dN>0$ as $dt>0$.

For $V=\lambda \phi^4/24$,  a value of $\lambda \sim 10^{-12}$ is required 
to get at least the normalization of the inflationary fluctuations right. 

Below we make use of the horizon-flow functions $\epsilon_n$ \cite{stg}, 
which are a 
generalization of the slow-roll parameters and are defined recursively: 
\begingroup
\everymath{\scriptstyle}
\small
\begin{eqnarray}
\label{psrpp} 
\epsilon_0 \equiv \frac{H_i}{H}, \quad 
\epsilon_{m+1}\equiv \frac{1}{\epsilon_m}\frac{d\epsilon_m}{dN},\,\,m \ge 0.
\end{eqnarray}
\endgroup
$H_i$ refers to the initial Hubble rate (at $N=0$). 
For single scalar field models of inflation, 
the first horizon flow function has the simple interpretation 
to be proportional to the ratio of the kinetic energy density to the 
total energy density of the field, 
\begingroup
\everymath{\scriptstyle}
\small
\begin{equation}
\epsilon_1 = 3 {\frac{\frac 12 \dot{\phi}^2}{ \frac 12 \dot{\phi}^2 + V}}.
\end{equation}
\endgroup
In the slow-roll approximation, the horizon-flow functions can be 
related to the usual slow-roll parameters \cite{lpb}.

\subsection{Initial Conditions}

In order to solve the horizon and flatness problems at or close to 
the Planck scale, we would need at least $50$ e-foldings. For 
the $\lambda \phi^4$ potential 
we can use the slow-roll approximation to estimate the amount of 
e-foldings to the end of 
inflation as $\Delta N \simeq (\phi^2 -\phi_e^2)/(8 M_p^2)$, with 
$\phi_e \simeq \sqrt{8} M_p$. Thus we need to specify the initial 
conditions of inflation at $\phi_i \geq 20 M_p$. 

In the spirit of chaotic inflation we would expect that 
$H_i \sim M_p, \dot{\phi} \sim M_p^2$ and $V \sim M_p^4$.  As mentioned
before, we explore here a situation where perturbations are evaluated just
before the onset of the slow-roll regime and thus must limit 
the potential to $V_i \sim M^4 \ll M_p^4$. This implies that 
$\epsilon_{1i} \approx 3$, since $V \ll \dot{\phi}^2$. Inflation 
requires $\epsilon_1 < 1$. This means that the Universe starts in a kinetic 
energy dominated regime. Due to the Hubble drag the kinetic energy density 
decays quickly and the dynamics becomes dominated by the 
potential energy density. Inflation starts and a slow-roll behaviour is 
approached.  
One has the freedom to choose  $0<\epsilon_{1i}<3$; in particular, 
$\epsilon_{1i} \simeq 2.9$ is 
compatible with chaotic inflation since the evolution of the system will drive 
it rapidly to the slow-roll attractor \cite{linde}. 

For the initial Hubble rate and the number of e-foldings, we use:
\begingroup
\everymath{\scriptstyle}
\small
\begin{eqnarray}
\label{eic}
H_i = \frac{1}{\sqrt{3-\epsilon_{1i}}}\sqrt{\frac{V_i}{M_p^2}}, 
\quad\quad
N_i = 0.
\end{eqnarray}
\endgroup

Starting from an initial condition as specified in the preceeding
paragraph and the use of the horizon flow functions, allow us to 
find different 
results from the ones already reported for the $\lambda\phi^4$ potential. 
The exclusion of $\lambda\phi^4$ inflation is
obtained in terms of parameters that measure the ratio of the potential to its 
derivatives, $\epsilon_{\rm V}\equiv M_p^2/2\left(V'/V\right)^2,\,\,
\eta_{\rm V}\equiv M_p^2V''/V$ \cite{lpb}. In doing  so, one implicitly 
assumes the validity of the slow-roll approximation and therefore 
omits the contribution of the kinetic energy before inflation starts, 
which in time excludes different possibilities for its set up. 

\begin{figure}
\includegraphics[width=85mm, height=55mm]{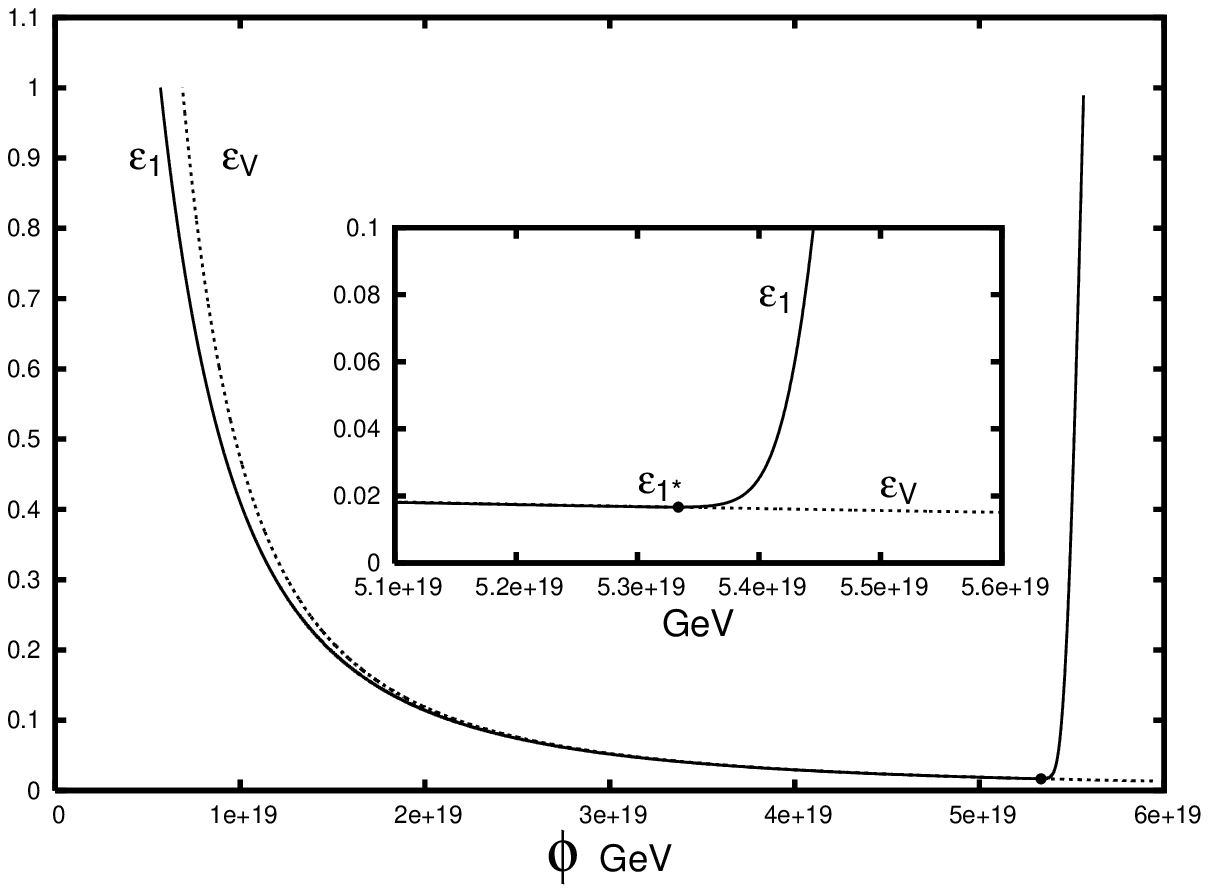}\\
\includegraphics[width=85mm, height=55mm]{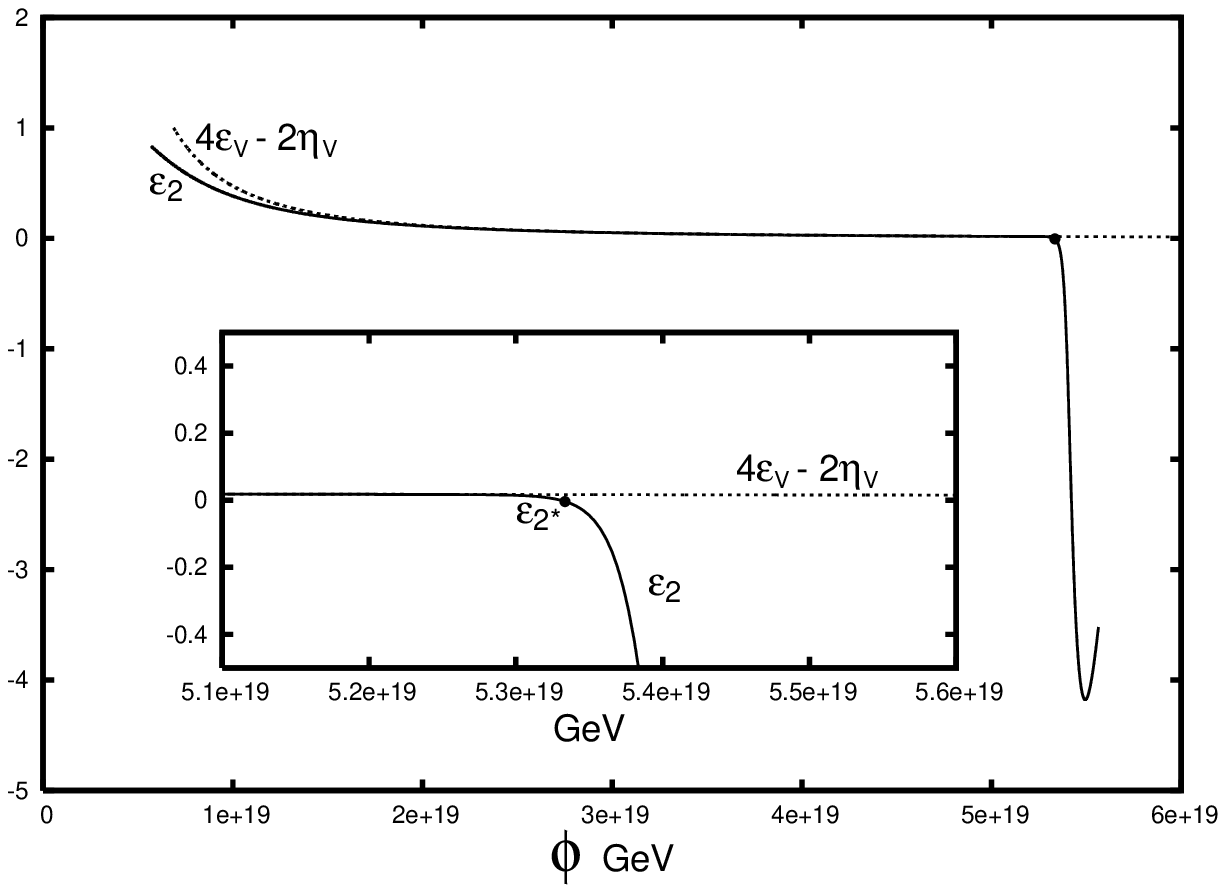}
\caption{Horizon flow functions and potential 
slow-roll parameters for an initial condition of $\epsilon_1=2.9$ and 
initial field value $\phi_i =24.4 M_p$. The $\bullet$ indicates 
the moment of time when modes observed in the CMB cross the horizon 
and where perturbations are evaluated. 
%Here we have used 
%$\epsilon_2=2(\epsilon_{\rm H}-\eta_{\rm H})$.
 %\cite{stg} and $\eta_{\rm H}\simeq \eta_{\rm V}-\epsilon_{\rm V}$
 %\cite{lpb}.
 }
\label{functions}
\end{figure}

In Figure \ref{functions} we can appreciate this difference; the 
region of slow-roll in 
both trajectories is the same, except that the horizon flow functions 
account for the full dynamics. The evolution of 
$\epsilon_1$ and $\epsilon_2$ agrees with that of $\epsilon_{\rm V}$ and 
$4\epsilon_{\rm V} - 2 \eta_{\rm V}$ within the slow-roll regime, but 
differs significantly 
at the onset of inflation and by a smaller amount towards the end.
 
\subsection{Spectrum of Density Perturbations}

An important condition that any inflationary model must fulfill 
is to produce the observed amplitude of the power spectrum of scalar 
perturbations
\begin{equation}
\label{amplitude} 
\Delta_R^2(k) \equiv \frac{k^3P_R(k)}{2\pi^2}.
\end{equation}
The best-fit amplitude at the pivot scale of $k_* = 0.002/$Mpc was reported as
$\Delta_R^2=(2.445\pm 0.096)\times 10^{-9}$ by the WMAP team 
\cite{wmap5}.
We follow the convention of WMAP5 and approximate the amplitude 
by a power-law:
\begingroup
\everymath{\scriptstyle}
\small
\begin{equation}
\label{p-l}
\Delta_R^2\left( k \right) \approx \left(\frac{H_*^2}
{8\pi^2 M_p^2\epsilon_{1*}}\right)A_*
\left(\frac{k}{k_*}\right)^{n_{s*}-1},
\end{equation}
\endgroup
where $n_s - 1$ is the spectral tilt. A $*$ indicates the scale at 
which perturbations are 
evaluated. $A$ denotes the first-order correction to the amplitude 
of the power spectrum in terms of the horizon-flow parameters \cite{llms}: 
\begin{equation}
A \approx  1- 2(C+1)\epsilon_{1} - C\epsilon_{2},
\end{equation}
where
$C\equiv \gamma_{\rm E}+\ln 2-2\simeq -0.7296$ and we have 
kept contributions of ${\cal O}(\epsilon_i)$ only . 

The approximation given by Eq.~(\ref{p-l}) does not consider the inclusion of 
running of the spectral index, in our results we obtain running 
and therefore make use of the power-law approximation for this situation 
given by:
\begingroup
\everymath{\scriptstyle}
\small
\begin{equation}
\label{p-l-r}
\Delta_R^2\left( k \right) \approx \left(\frac{H_*^2}
{8\pi^2 M_p^2\epsilon_{1*}}\right)A_*
\left(\frac{k}{k_*}\right)^{n_{s*}-1
+\frac{1}{2}\frac{dn_{s*}}{d\ln\left({k}/{k_*}\right)}
\ln\left({k}/{k_*}\right)},
\end{equation}
\endgroup
where $\frac{dn_s}{d\ln(k/k_*)}$ is the running of the spectral index.

The tensor-to-scalar ratio $r$ and the spectral tilt are at leading order in 
the horizon-flow functions given by:
\begingroup
\everymath{\scriptstyle}
\small
\begin{eqnarray}
\label{ob}
r&\approx& 16\epsilon_{\rm 1},\\ \nonumber
n_{\rm s} - 1 &\approx& -2\epsilon_{\rm 1}-\epsilon_{\rm 2}.
\end{eqnarray}
\endgroup 

\subsection{End of Inflation and Pivot Scale}

The end of inflation is given by the condition $\epsilon_1=1$ that
fixes the total number of e-foldings $N_e$, and the value of the field 
at the end of inflation $\phi_e$. Let  $\Delta N_*\equiv N_e-N_*$ be 
the number of e-foldings between horizon crossing of the 
pivot scale $k_*$ and the end of inflation. 

For the comparison of the theoretical power spectra with 
observations, we choose values of $\Delta N_*$ within 
a certain interval. 
Then we calculate the pivot scale $k_* = a_* H_*$. In order to do this, 
one needs a model 
of reheating, and for the moment we assume sudden reheating to a 
radiation-dominated 
Universe at the end of inflation, which means that our values 
are upper estimates for $\Delta N_*$. 

We know that, 
\begingroup
\everymath{\scriptstyle}
\small 
\begin{eqnarray}
\epsilon_1=1\Rightarrow \dot{\phi}^2_{e}=V_{e}, \quad
H^2_{e}=\frac{1}{2M_p^2}V_{e}=\frac{1}{3M_p^2}\rho_{r} 
\end{eqnarray}
\endgroup
with $\rho_{r}=(\pi^2/30) g T^4_{r}$ and 
$T_{r}^4=\frac{45}{\pi^2}\frac{V_e}{g}$ being the energy density and
temperature at reheating. $g$ 
is the effective number of relativistic helicity degrees of freedom. Above 
$T \sim 100$ GeV
we take the value $g =  106.75$. Therefore, for the pivot scale 
$k_* = a_*H_*$ we have from $k(\phi)/k_* =  e^{\Delta N}H(\phi)/H_*$,
where $\Delta N = N(\phi)-N_*$ 
and with $a_0 = 1$:
\begingroup
\everymath{\scriptstyle}
\small 
\begin{eqnarray}
k_*&=& e^{-\Delta N_*}H_*\frac{a_e}{a_0}a_0 
= e^{-\Delta N_*}H_* (1+z)_e^{-1} \\ \nonumber
&=& 500e^{-\Delta N_*}H_*\frac{T_{\nu_0}}{T_{reh}}
\left(0.002 {\rm Mpc}^{-1}\right),
\end{eqnarray} 
\endgroup
where the redshift to the end of inflation 
is given by the ratio of today's neutrino temperature to the 
reheating temperature.

\subsection{Methodology}

Our scenario of inflation is controlled by three parameters: the
initial value of the inflaton field,   
$\phi_i ={\tt a}\,M_p$, the initial value of $\epsilon_1$ and the scale at 
which one evaluates the perturbations, namely, the value of $\Delta N_*$. 
This last point has to be considered carefully since an initial value of 
$\epsilon_{1i} >1$ 
gives less inflation on the whole trajectory than a value of 
$\epsilon_{1i} \approx 0$. 
Additionally, we want to evaluate perturbations at the point when
$|\epsilon_2|\ll 1$ and $\epsilon_2<0$. The reason being that 
the case $0<\epsilon_2<1$ gives the values already excluded by the 
WMAP plus baryonic acoustic oscillations plus super novae
(WMAP+BAO+SN) analysis. A $\Delta N_*$ lying on 
the region where both trajectories in Fig.~\ref{functions} are the same, 
will resort to values of $r$ and $n_s$ already excluded. 

\section{Results}

For our results we set $\lambda$ to $10^{-12}$.
The values for ${\tt a} = \phi_i /M_p$, 
$\epsilon_1$ and $\Delta N_*$ are chosen 
within the following intervals:
\begingroup
\everymath{\scriptstyle}
\small
\begin{eqnarray}
\label{init_cond1}
\epsilon_1 \in [0.5,2.9], \quad \Delta N_* \in [58,65], \quad {\tt a} 
\in [20,35].
\end{eqnarray}
\endgroup
The interval of values for ${\tt a}$ assures that the value of $\phi_{i}$ is 
well above $3 M_p$ (the end of inflation) and allows for at least $50$ 
e-foldings, corresponding to 
the lower limit of the interval. 
The upper limit takes into account that during the onset of inflation 
the short epoch of fast 
roll does not give rise to an exponential expansion of the Universe.

\begin{figure}
\includegraphics[width=85mm, height=55mm]{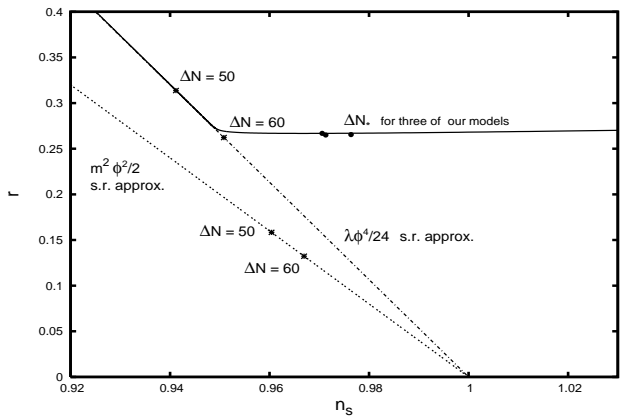}
\caption{ Tensor-to-scalar ratio $r$ versus the spectral index $n_{\rm s}$ 
for the 
three models detailed in Table~\ref{ch_inf} and the trajectory of the 
second of them. 
Our examples violate the slow-roll approximation. For comparison the 
slow-roll 
trajectories of the $\lambda \phi^4$ and the $m^2 \phi^2$ potentials are 
shown.}
\label{observ_1}
\end{figure}

The location in parameter space for three cases are shown in
Figure~\ref{observ_1} and their predictions given explicitly in
Table~\ref{ch_inf}.  
From the figure, it is possible to appreciate that the values 
for the tensor-to-scalar ratio $r$ are approximately the same as 
in the usual slow-roll approximation, since the interval of values 
of $\Delta N_*$ is concentrated around 60 e-foldings. The crucial 
difference is given by the value of $\epsilon_{2*}$, which at this 
level of approximation enters only in the spectral tilt and 'shifts' 
the points towards the right. These results cannot be obtained in a 
model that considers initial conditions for the system in the slow-roll 
regime. Our results for the spectral index are
consistent with the bounds of WMAP5 for Running+Tensors at the $2\sigma$
level, the running and the tensor-to-scalar ratio as well as the amplitude of
the spectrum are inside the $1\sigma$ interval.

\begin{table*}
\caption{Examples of ``just enough'' chaotic inflation for 
$V = \lambda \phi^4/24$ and $\lambda = 10^{-12}$.}
\begin{tabular}{|c|c|c|c|c|c|c|c|c|c|c|c|c|}\hline
$\epsilon_{1,i}$ &
$\phi_i/M_p$ &
$V_i/M_p^4$ &
$N_T$ &
$\Delta N_*$ & 
$\epsilon_{1*}$ &
$\epsilon_{2*}$ &
$\epsilon_{3*}$ &
$r_*$ &
$n_{s*}$ &
$dn_{s*}/{dln k}$ &
${H_*^2 A}/(8\pi^2 M_p^2\epsilon_{1*})$ &
$k_*$ \\ \hline
    
$0.77$ &
$22.832$ &
$1.1\times 10^{-8}$ &
$62.96$ &
$60.50$ &
$1.659\times 10^{-2}$ &    
$-9.555\times 10^{-3}$ &
$-8.176$ &
$0.27$ &
$0.98$ &
$-7.78\times 10^{-2}$ &
$2.45 \times 10^{-9}$ &
$0.010$ \\ \hline

$2.9$  &
$24.4$ &
$1.1\times 10^{-8}$ &
$62.69$ &
$60.08$ &
$1.668\times 10^{-2}$ &  
$-3.939\times 10^{-3}$ &
$-15.629$ &
$0.27$ &
$0.97$ &
$-6.143\times 10^{-2}$ &
$2.42\times 10^{-9}$ &
$0.015$ \\ \hline

$2.2$ &
$23.5$ &
$1.2\times 10^{-8}$ &
$63.10$ &
$60.50$ &
$1.656\times 10^{-2}$ &   
$-4.42\times 10^{-3}$ &
$-14.178$  &
$0.27$ &
$0.97$ &
$-6.25\times 10^{-2}$ &
$2.47 \times 10^{-9}$ &
$0.010$ \\ \hline
\end{tabular}
\label{ch_inf}
\end{table*}

\begin{figure}
\includegraphics[width=85mm, height=55mm]{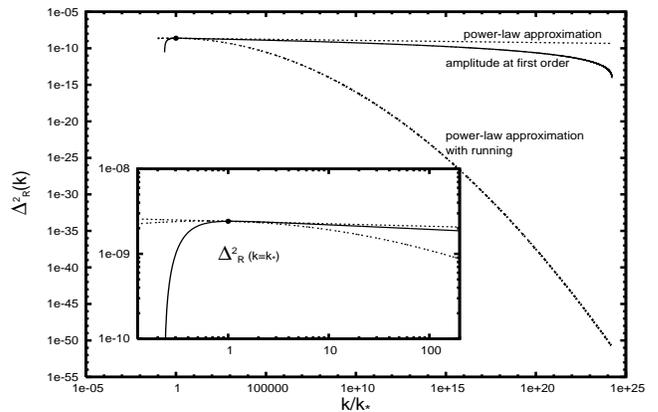}
\caption{Power spectrum of scalar fluctuations at first order in the 
horizon flow functions, its power-law approximation without running, and the
power-law approximation including running. The $\bullet$ indicates 
where perturbations are evaluated.}
\label{asr}
\end{figure}

For the second of the examples shown in Table~\ref{ch_inf} 
the spectrum produced and the approximation given by Eq.~(\ref{p-l}) and
 Eq.~(\ref{p-l-r})
are presented in Fig.~\ref{asr}. The zoom in the figure gives 
an idea about how the 
spectrum looks in the actually observed region of the power spectrum.

\begin{figure}
\includegraphics[width=85mm, height=55mm]{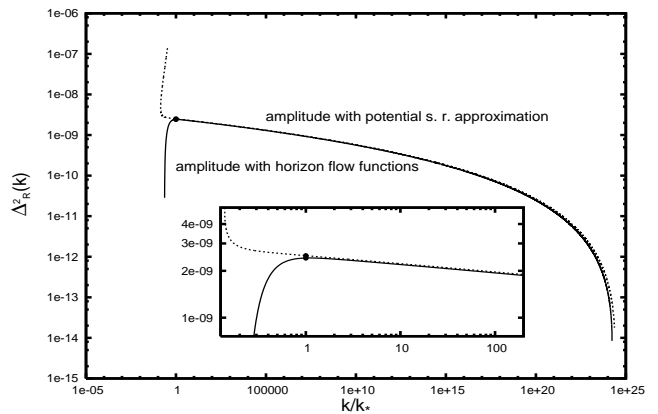}
\caption{Power-spectrum of scalar fluctuations: the upper curve 
corresponds to
using $\epsilon_{\rm V},\,\eta_{\rm V}$ and the leading slow-roll 
approximation. 
The lower-curve is the result of using $\epsilon_1,\,\epsilon_2$. 
The dots indicate where perturbations are evaluated.}
\label{hsr}
\end{figure}

The difference between the spectra produced using $\epsilon_1,\,\,\epsilon_2$
and $\epsilon_{\rm V},\,\,\eta_{\rm V}$ can be appreciated in Fig.~\ref{hsr},
the region shown corresponds to the onset of the slow-roll regime during which
the behavior of both is indistinguishable. Before the point where fluctuations
are evaluated at $k < k_*$, there is however, considerable difference 
between both of them as a consequence of the dynamics. 

\begin{figure}
\includegraphics[width=85mm, height=55mm]{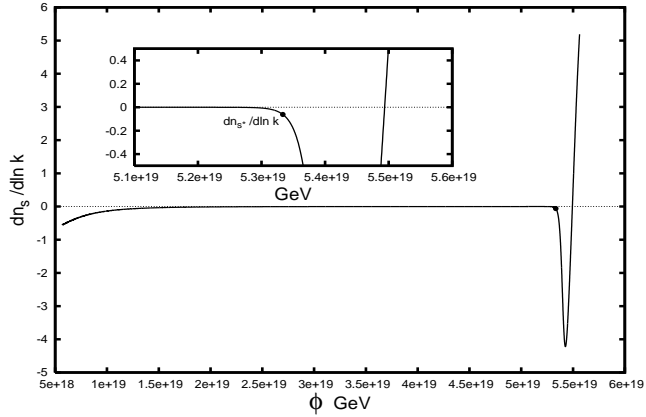}
\caption{Running of the spectral index as a function of the inflaton field
for $\epsilon_{1i} = 2.9$ and $\phi_i =24.4$. The $\bullet$ 
indicates the moment of time when the pivot wave number $k_*$ crosses the 
horizon 
and where perturbations are evaluated.}
\label{runing}
\end{figure}

In Fig.~\ref{runing}, the running of the spectral index,
\begingroup
\everymath{\scriptstyle}
\small 
\begin{equation}
\frac{dn_s}{d\ln(k/k_*)}\approx
-\epsilon_2\left(2\epsilon_1+\epsilon_3\right)
\end{equation}
\endgroup
shows that
perturbations are evaluated when the system has not yet arrived to the region
when the spectrum can be approximated completely by a power-law. The value of
the running is negative for all cases and of order  
${\cal O}(10^{-2})$. This can be explained by noting that at 
the point where perturbations are evaluated, $\epsilon_2$ has not yet
arrived to the slow-roll regime \cite{stg,ste}. In fact it is going to 
change sign and thus higher-order horizon flow functions will diverge at that 
point, which just means that the slow-roll approximation cannot be applied. 
Strictly speaking we should solve the mode equations for the perturbations 
numerically to obtain a more reliable result. However, it seems to us that 
for a first study of our scenario the current level of sophistication is 
good enough. 

The values of the scale $k_*$ at which perturbations are evaluated for our
results are bigger by a factor of 5 than the pivot scale used by WMAP5. One
has to bear in mind the uncertainty due to assuming sudden reheating for  
obtaining these numbers.

The initial value of the potential 
at the Planck scale is determined by the quartic coupling $\lambda$ and
$\phi_i$, these two numbers set the potential 
to 8 orders of magnitude below $M_p^4$. This confirms the scenario 
described above, which 
differs from the usual ``chaotic'' initial conditions where 
kinetic and potential energies are of the same order of magnitude initially. 
We have seen that the initial condition of $\epsilon_1$ very close 
to 3 is needed
in order to also have $H_i$ of the order of the Planck mass. This initial 
condition has also allowed us to obtain acceptable values for the 
inflationary observables.  

However, from Table~\ref{ch_inf} it is possible to see that an initial 
$H_i\sim M_p$ is not a necessary requirement to ensure acceptable 
$r$ and $n_s$; an initial value as low as $0.77$ for $\epsilon_1$ gives also
acceptable models. Therefore, an initial $\epsilon_1$ 
that gives convenient values for the observables need not be arbitrarily 
close to 3, but only outside of the slow-roll regime. Nevertheless, in our 
scenario a value of $\epsilon_{1i} \approx 3$, seems most natural to us.

\section{conclusions}

From the results shown in Figure~\ref{observ_1} 
and Table~\ref{ch_inf}, we conclude that the $\lambda\phi^4$ inflaton 
potential is not excluded. This conclusion is obtained within a set up that
allows for the existence of a fundamental maximal scale of the inflaton 
potential $V \sim M^4$, which is well below the Planck scale $M_p^4$ at the 
beginning of inflation. Such a scale arises naturally in the context of
the standard model of particle physics and some of its suggested extensions. 
Loop corrections to the effective potential might become imaginary, leading to 
a destabilization of the inflaton field. In supergravity theories of 
inflation the 
potential might become too steep. Both instances would define a scale 
$M < M_p$. Below $M$ all interactions apart from gravity are taken into 
account by the effective potential. From the results of Table~\ref{ch_inf}, 
it is consistent to assume $M$ to be the scale of grand unification 
(GUT scale), which is at least a pleasing coincidence. 
 
The use of the horizon flow functions instead of the potential slow-roll 
parameters allowed us to follow the cosmological evolution from the onset 
of inflation for the model considered here. The difference between both
approaches has also been shown in more complicated models of inflation 
like the hybrid inflation scenario \cite{CR}.

As is obvious 
from Fig.~\ref{hsr}, the power-law approximation either with running or
without it, 
is not describing the actual power spectrum very well. Both of the 
approximations overestimate the power at large length scales, the
approximation without running is closer to the true amplitude on small scales. 
The approximation with running significantly underestimates the power at small 
scales.
This implies that for a detailed analysis of the epoch of the onset of 
slow-roll, 
we cannot just take the published fits from the WMAP analysis on 
$\Delta_R^2$, $n_s$ 
and $r$, but we should
run a new Markov chain Monte Carlo integration. The fact that we find values 
of correct order of 
magnitude makes us confident that a more refined analysis will also 
allow us to find a good fit. 
The calculation of the power spectrum itself should be based on 
a numerical mode-by-mode 
integration, as the expansion in horizon flow functions that we are 
using here is at best an estimate 
of the true power spectrum. The usual slow-roll approximation, as 
argued above, is even worse.  

As already noted, we find a power spectrum that is not featureless. The 
amount of running is in agreement with observations. On top of that, we
find a suppression of power on the largest scales. 
This suppression could be linked to the lack of CMB correlations on 
large angular scales \cite{Spergel,Copi}.
In order to explain the lack
of power on large scales it was proposed before to consider a 
period of fast-roll at the beginning of inflation and to have
only 60 to 65 e-foldings of expansion \cite{Contaldi,Nicholson} (very
much like in our scenario), or to assume a fast-roll epoch in between two 
epochs of slow-roll \cite{Jain}. The difference to our work is however,
that different inflationary potentials have been used and it seems to us
that our set-up is more natural. Apart from the lack of large scale
correlation,  
the observed alignment of quadrupole and octupole \cite{deOliveira,Copi2} 
could be a remainder of the not yet perfect statistical isotropy at the onset 
of inflation. A detailed study of those aspects is beyond the scope 
of this work.   

This alternative set up for inflation, based on the
existence of a second fundamental scale $M < M_p$, leads us to the
consideration of models that are not properly described by slow-roll
inflation (but nevertheless contain an epoch of slow-roll).  
To exclude a specific model of inflation by means of observations, one needs 
to test whether the fluctuations are evaluated in the slow-roll regime.
Improved analysis of available and upcoming data from WMAP and from
the Planck satellite will allow us to probe the new scenario of 
"just enough" chaotic inflation.

\section{acknowledgments}

We thank S.~Clesse, M.~Einhorn, G. Isidori, T.~Jones, G.~Moore, H.~Solis, 
L.~Sriramkumar A. Strumia and T.~Takahashi for interesting discussions 
and/or hints to the literature. E.~R.~was supported by DAAD research 
grant A/07/26003 and Conacyt grant 79045.

\end{document}